\documentclass[preprint2]{aastex63}

\def\gsim{\mathrel{\rlap{\lower 4pt \hbox{\hskip 1pt $\sim$}}\raise 1pt
\hbox {$>$}}}
\def\lsim{\mathrel{\rlap{\lower 4pt \hbox{\hskip 1pt $\sim$}}\raise 1pt
\hbox {$<$}}}

\received{}
\revised{}
\accepted{}
\shorttitle{CSM around SN Ic 2020oi}
\shortauthors{Maeda et al.}
\graphicspath{{./}{figures/}}

\begin{document}

\title{The final months of massive star evolution from the circumstellar environment around SN Ic 2020oi}

\correspondingauthor{Keiichi Maeda}
\email{keiichi.maeda@kusastro.kyoto-u.ac.jp}

\author[0000-0003-2611-7269]{Keiichi Maeda}
\affiliation{Department of Astronomy, Kyoto University, Kitashirakawa-Oiwake-cho, Sakyo-ku, Kyoto, 606-8502. Japan}

\author[0000-0002-0844-6563]{Poonam Chandra}
\affiliation{National Centre for Radio Astrophysics, Tata Institute of Fundamental Research, Ganeshkhind, Pune 411007, India}

\author[0000-0002-6916-3559]{Tomoki Matsuoka}
\affiliation{Department of Astronomy, Kyoto University, Kitashirakawa-Oiwake-cho, Sakyo-ku, Kyoto, 606-8502. Japan}

\author[0000-0003-4501-8100]{Stuart Ryder}
\affiliation{Department of Physics and Astronomy, Macquarie University, NSW 2109, Australia}
\affiliation{Macquarie University Research Centre for Astronomy, Astrophysics \& Astrophotonics, Sydney, NSW 2109, Australia}

\author[0000-0003-1169-1954]{Takashi J. Moriya}
\affiliation{National Astronomical Observatory of Japan, National Institutes of Natural Sciences, 2-21-1 Osawa, Mitaka, Tokyo 181-8588, Japan}
\affiliation{School of Physics and Astronomy, Faculty of Science, Monash University, Clayton, Victoria 3800, Australia}

\author[0000-0002-1132-1366]{Hanindyo Kuncarayakti}
\affiliation{Tuorla Observatory, Department of Physics and Astronomy, FI-20014 University of Turku, Finland} 
\affiliation{Finnish Centre for Astronomy with ESO (FINCA), FI-20014 University of Turku, Finland}

\author[0000-0002-2899-4241]{Shiu-Hang Lee}
\affiliation{Department of Astronomy, Kyoto University, Kitashirakawa-Oiwake-cho, Sakyo-ku, Kyoto, 606-8502. Japan}

\author[0000-0002-4807-379X]{Esha Kundu}
\affiliation{International Centre for Radio Astronomy Research, Curtin University, Bentley, WA 6102, Australia}

\author[0000-0002-7507-8115]{Daniel Patnaude}
\affiliation{Smithsonian Astrophysical Observatory, Cambridge, MA 02138, USA}

\author{Tomoki Saito}
\affiliation{Nishi-Harima Astronomical Observatory, Center for Astronomy, University of Hyogo, 407-2 Nishigaichi, Sayo, Sayo, Hyogo 679-5313, Japan}

\author[0000-0001-5247-1486]{Gaston Folatelli}
\affiliation{Instituto de Astrof{\'i}sica de La Plata (IALP), CONICET, Argentina}
\affiliation{Facultad de Ciencias Astron{\'o}mcas y Geof{\'i}sicas, Universidad Nacional de La Plata, Paseo del Bosque, B1900FWA, La Plata, Argentina}
\affiliation{Kavli Institute for the Physics and Mathematics of the Universe (WPI), The University of Tokyo, Institutes for Advanced Study, The University of Tokyo, 5-1-5 Kashiwanoha, Kashiwa, Chiba 277-8583, Japan}



\begin{abstract}
We present the results of ALMA band 3 observations of a nearby type Ic supernova (SN) 2020oi. Under the standard assumptions on the SN-circumstellar medium (CSM) interaction and the synchrotron emission, the data indicate that the CSM structure deviates from a smooth distribution expected from the steady-state mass loss in the very vicinity of the SN ($\lsim 10^{15}$ cm), which is then connected to the outer smooth distribution ($\gsim 10^{16}$ cm). This structure is further confirmed through the light curve modeling of the whole radio data set as combined with data at lower frequency previously reported. Being an explosion of a bare carbon-oxygen (C+O) star having a fast wind, we can trace the mass-loss history of the progenitor of SN 2020oi in the final year. The inferred non-smooth CSM distribution corresponds to fluctuations on the sub-year time scale in the mass-loss history toward the SN explosion. Our finding suggests that the pre-SN activity is likely driven by the accelerated change in the nuclear burning stage in the last moments just before the massive star's demise. The structure of the CSM derived in this study is beyond the applicability of the other methods at optical wavelengths, highlighting an importance and uniqueness of quick follow-up observations of SNe by ALMA and other radio facilities. 
\end{abstract}



\section{Introduction} \label{sec:intro}

The core-collapse supernova (CCSN) is an explosion of a massive star following the exhaustion of nuclear fuel and the subsequent core collapse \citep{langer2012}. An increasing opportunity of early discovery of new SNe and quick follow-up observations at optical wavelengths has opened up a new window to study the nature of the circumstellar medium (CSM) in the vicinity of SNe, which is then translated to the nature of the mass-loss history and pre-SN activity just before the explosion. This investigation has revealed that a large fraction of SNe II have a dense CSM which extends up to a few $\times 10^{15}$ cm, as is frequently termed the `confined CSM' \citep{gal-yam2014,khazov2016,yaron2017,forster2018}. If the mass-loss velocity is $v_{\rm w} \sim 10$ km s$^{-1}$ for the extended progenitors of SNe II (mainly red supergiants; RSGs) \citep{smith2014a,moriya2017}, this confined CSM must have been created by the pre-SN activity in the last $\sim 30$ yrs, with the corresponding mass-loss rate of $\sim 10^{-3} M_\odot$ yr$^{-1}$ \citep{groh2014,morozova2015,moriya2017,yaron2017,forster2018}. This is much larger than the usual mass-loss rate \citep{smith2014a} derived from the outer CSM distribution ($\sim 10^{-7} - 10^{-8} M_\odot$ yr $^{-1}$) \citep{yaron2017}. 

As another piece of evidence for the pre-SN activity, detection of pre-SN outbursts has been reported \citep{pastorello2007,ofek2013,ofek2014,smith2014c,strotjohann2021} for rare classes of CCSNe \citep{li2011}, i.e., SNe IIn and Ibn, showing strong signatures of the SN-CSM interaction in the optical. However, the nature of their progenitor stars has not been well determined \citep{moriya2014,moriya2016}. It is not clear if the pre-SN activity observed for these SNe is representative of the massive star evolution. 

The mechanism leading to the pre-SN activity in the end of the stellar life has not been clarified. A popular suggestion is that this may be related to the rapidly increasing energy generation by progressively more advanced nuclear burning stages. This final phase may not be represented by a classical `static' stellar evolution theory \citep{arnett2011,smith2014b}, which might underestimate the nuclear energy generation. Further, the generated energy in the core can exceed the hydrostatic limit, and it may be tunneled toward the envelope as a wave \citep{quataett2012,fuller2017}. The envelope may then dynamically respond to the core evolution \citep{ouchi2019,morozova2020}. The rapid core evolution may also be coupled with the envelope through the angular momentum transport and could induce the pre-SN mass loss \citep{aguielra2018}. All these possible processes have not been taken into account in the classical stellar evolution theory. 

The core evolution is accelerated toward the formation of the iron core. The mass-loss history in the last $\sim 10-100$ years previously investigated for SNe II corresponds to the carbon burning stage, which lasts from $\sim 1,000$ yrs to $\sim 10$ years before the SN explosion \citep{langer2012,fuller2017}. To understand the origin of the pre-SN final activity, one wants to go much closer to the end of the stellar life; neon burning commences only a few years before the explosion, and oxygen burning is activated in the final one year. 

The so-called stripped envelope SNe (SESNe) \citep{filippenko1997} provide a good opportunity here. SESNe include SNe Ib from a He star progenitor and SNe Ic from a C+O star progenitor \citep{langer2012}. The typical mass-loss wind velocity of the SESN progenitors is $v_{\rm w} \sim 1,000$ km s$^{-1}$ \citep{chevalier2006,crowther2007,smith2014a}; CSM at $\sim 10^{15}$ cm must then have been ejected by a progenitor star at $\sim 0.3$ year before the explosion. Given the high wind velocity, the expected CSM density would not be sufficiently high to leave a strong trace in the optical (which will be further discussed in the present paper, Section 5). 

Indeed, the `flash' spectroscopy within a few days has been mostly limited to SNe II \citep{shivvers2015,khazov2016,yaron2017,bruch2020}. An exception is SN IIb 2013cu \citep{gal-yam2014}, which represents a transitional object between SNe II and SNe Ib/c. However, the analyses of its flash spectra indicate that the mass-loss velocity of the progenitor is $v_{\rm w} \lsim 100$ km s$^{-1}$ and most likely $\sim 30$ km s$^{-1}$ \citep{groh2014,grafener2016}. The progenitor is thus more consistent with the one with an extended H-rich envelope, rather than a genuine He star, being similar to SN IIb 2011dh as a representative case \citep{maund2011,bersten2012}. The other exceptions are SN Ic 2014ft for which a dense and confined H-poor CSM was inferred from a flash spectrum \citep{de2018}, and broad-lined SN Ic 2018gep for which the pre-SN activity was inferred from its precursor emission \citep{ho2019}. SNe Ic 2014ft and 2018gep are however peculiar outliers and their progenitor evolution is unlikely to be representative of a bulk of SESNe. Furthermore, the evolution of the mass-loss rate in the final few years has not been quantified in detail for these SESNe. 

A few examples exist, e.g. SN Ib 2004dk \citep{pooley2019,bal2021}, SN Ib 2014C \citep{anderson2017,margutti2017,tiny2019}, SN Ic 2017dio \citep{kuncarayakti2018}, and SN Ib 2019oys \citep{sollerman2020}, where the ejecta of a C+O/He progenitor interact with dense CSM to produce strong emissions either in the optical or in the radio, or both. However, the dense CSM in these SNe was found to be located at $\gsim 10^{16}$ cm. Such relatively distant CSM should not be created by the pre-SN activity in the last few years. Indeed, they might reflect a rare channel in the binary evolution \citep{ouchi2017,kuncarayakti2018}, while the bulk of SESNe are thought to experience the binary interaction at much earlier times \citep{yoon2017,fang2019}. It is necessary to trace the CSM distribution of SNe Ib/c at $\lsim 10^{15}$ cm to probe the evolution of a massive star in the last few years, which has however been challenging so far at optical wavelengths. The nature of CSM within $\sim 10^{15}$ cm has been largely unexplored for SNe Ib/c. 

This makes radio observation a unique tool, through the synchrotron emission exclusively created by the SN-CSM interaction \citep{bjornsson2004,chevalier2006,maeda2012,matsuoka2019,horesh2020}. Multi-band radio observations for SESNe in the infant phase ($\lsim 10$ days) have however been very limited, suffering from a lack of wavelength or temporal coverage, especially in the high frequency \citep{berger2002,weiler2002,soderberg2010,soderberg2012,horesh2013a,horesh2013b,kamble2016,bietenholz2021}. The situation is similar for broad-lined SNe Ic, for which radio follow-up observation is routinely undertaken \citep[e.g., ][]{corsi2016}. They tend to show the CSM density at $10^{16}$ cm being lower than for typical SESNe except for a few cases \citep{tetteran2019,nayana2020}, while little is known about the nature of CSM at the scale of $\sim 10^{15}$ cm. While very rapid radio follow-up observations have been conducted for a few broad-lined SNe Ic associated with a long Gamma-Ray Burst (GRB), the physical scale of the CSM probed at a few days after the explosion is already at $\gsim$ a few $\times 10^{15}$ cm for these GRB-SNe due to the (sub) relativistic ejecta creating the synchrotron emission \citep{kulkarni998}. 

The best observed case among SESNe so far would be SN IIb 2011dh \citep{horesh2013a}, but its progenitor has been derived to be an extended star \citep{maund2011} with low $v_{\rm w}$. Another good example of quick radio follow-up observation is SN Ib iPTF13bvn, whose progenitor is probably a compact He star \citep{cao2013,folatelli2016}. The radio data, including that at 100 GHz, were however still sparse and limited to the earliest phase (up to $\sim 10$ days) \citep{cao2013}, which is not sufficient to characterize the CSM distribution at different scales. 

The recent development of high-cadence surveys now allows multi-wavelength follow-up observations of SNe in the infant phase \citep{bellm2019,graham2019}. SN Ic 2020oi in the nearby galaxy M100 ($\sim 15$ Mpc) was discovered on 7 Jan 2020, 13:00:54 (UT) in a very infant phase ($\sim 1$ day after the putative explosion date; 6 Jan 2020, JD $58854.0 \pm 1.5$) \citep{forster2020,horesh2020,siebert2020,rho2021}. In this paper, we present the data from our ALMA observations for SN 2020oi (Section 2). We then investigate the nature of the CSM surrounding SN 2020oi using the ALMA band 3 data (at 100 GHz) as combined with the lower frequency observations presented by \citet{horesh2020} from 5 GHz to 44 GHz (Section 3). Based on the analyses in Section 3, we further perform detailed light curve model calculations in Section 4. The results of Sections 3 and 4 show that the CSM structure around SN 2020oi deviates from a single-power law distribution, indicating that the mass-loss characteristics show fluctuations on the sub-year time scale toward the SN explosion, as discussed in Section 5. Discussion in Section 5 further includes possible limitations in the present work, other possible explanations, and further details on the treatment of physical processes involved in the interpretation. The paper is closed in Section 6 with a summary.

\section{Observations and Data Reduction}\label{sec:obs}

\begin{deluxetable}{lll}
\tablenum{1}
\tablecaption{ALMA band 3 measurement of SN 2020oi (100GHz)\label{tab:flux}}
\tablewidth{0pt}
\tablehead{
\colhead{MJD} & \colhead{Phase} & \colhead{$F_{\nu}$ (with $1\sigma$ error)} \\
\colhead{} & \colhead{(Days)} & \colhead{(mJy / beam)} 
}
\startdata
58859.4 &  5.4 & $1.300 \pm 0.190$\\
58862.4 & 8.4 & $1.219 \pm 0.084$\\
58872.3 & 18.3 & $0.196 \pm 0.058$\\
58905.4 & 51.3 & $0.115 \pm 0.043$ \\
\enddata
\tablecomments{The phase is measured from the putative explosion date (MJD 58854.0) \citep{horesh2020}.
}
\label{tab;flux}
\end{deluxetable}

\begin{figure*}[t]
\centering
\includegraphics[width=1.5\columnwidth]{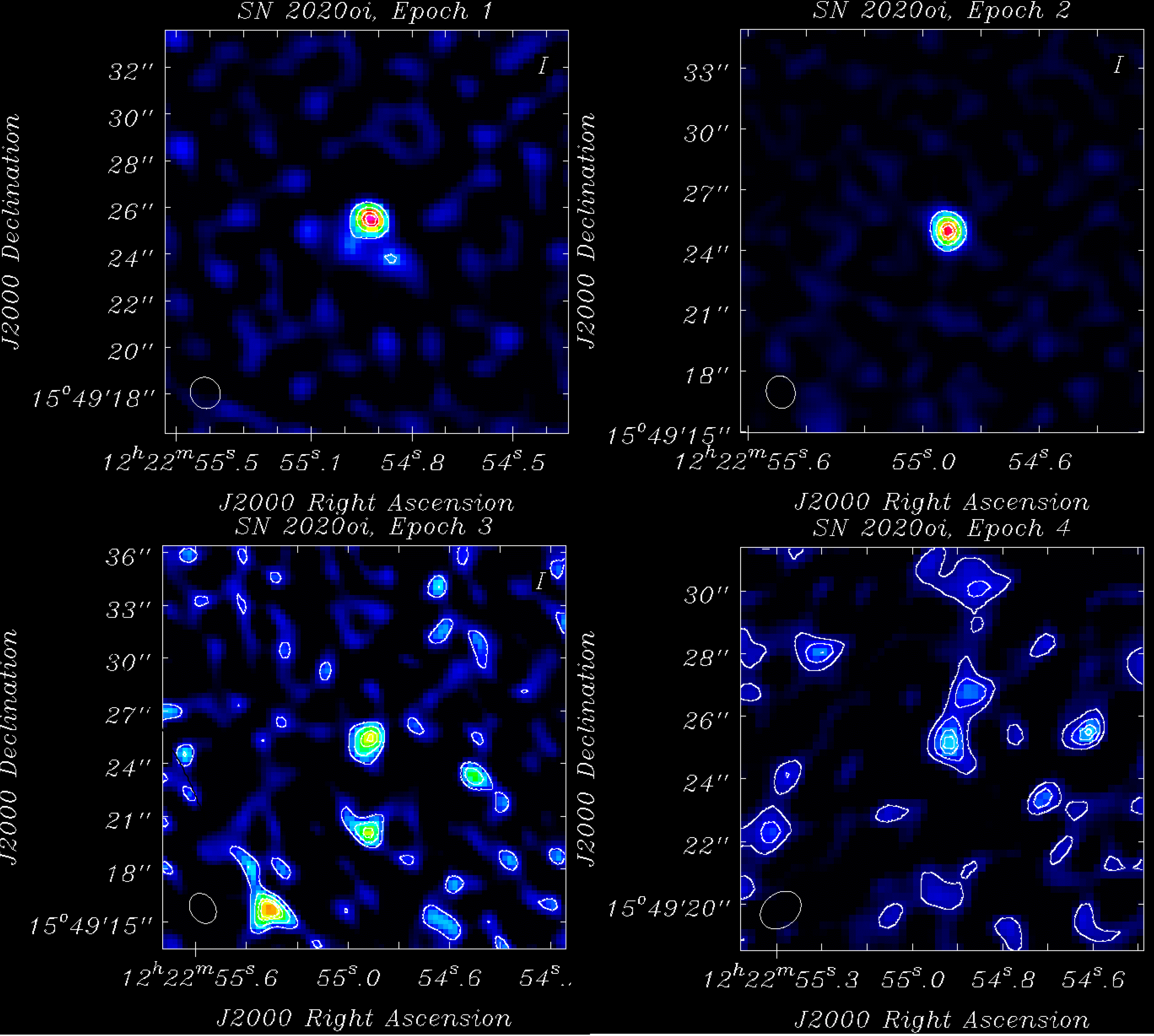}
\caption{The ALMA band 3 (the central frequency at 100 GHz) images of SN 2020oi, at 5.4 (top-left), 8.4 (top-right), 18.3 (bottom-left), and 51.2 (bottom-right) days since the putative explosion date. The color is normalized by the flux density range [0.0 mJy:1.5 mJy] for the earlier two epochs and [0.0 mJy: 0.3 mJy] for the later two epochs. The contours represent 35, 60, 80, and 90\% of the peak flux density. The elliptical beam shape is shown on the left-bottom corner in each panel. 
}
\label{fig:image}
\end{figure*}

Our ALMA Target-of-Opportunity (ToO) observations, as a part of cycle 7 high-priority program 2019.1.00350.T (PI: KM), have been conducted starting on 11 Jan 2020 (UT) covering 4 epochs. The log of the ALMA observations is shown in Tab. \ref{tab:flux}. The on-source exposure time is $16-20$ min per epoch. All the observations were conducted with band 3, with the same spectral set up for all the observations; the central frequency is 100 GHz, composed of 4 single continuum windows with the band width of 2 GHz each. Potentially strong molecular bands were avoided in setting the spectral windows, e.g., CO ($J=0-1$). The arrays are in the C-3 configuration, with the baselines $\sim 15-500$m or $\sim 15-783$m. After the image reconstruction as described below, the angular resolutions are $\sim 1.2-1.6$" which guarantee the minimum contamination by known sources, including the core of M100; the SN is located 1.3" east and 6.5" north of the center of M100. 

The data have been calibrated through the standard ALMA pipeline with CASA version 5.6.1-8. The image reconstruction has been done with additional manual processes with the CASA $tclean$ command. We have used the Briggs scheme with the Robustness parameter between $-0.5$ and $0.5$ in weighting the visibility prior to imaging. We have also introduced a minimum baseline cut up to $40k \lambda$ to assure no contamination from possible diffuse sources. The final measurements have then been performed by the CASA $imfit$ command, which provide the consistent values between the integrated source flux density and peak flux density per beam as expected for a point source. The final error includes the error in $imfit$, image rms, and the error in flux calibration. The flux densities are reported in Tab. \ref{tab:flux}. The reconstructed images are shown in Fig. \ref{fig:image}. A radio point source is clearly detected. The source keeps fading during our observations, and thus it is robustly identified as SN 2020oi.

The lower frequency, excellent data set at 5-44 GHz are taken from \citet{horesh2020}, which were obtained by ATCA, VLA, AMI-LA, and e-MERLIN. In our analysis, we omit the data for which they suspect nearby source contamination. The multi-band light curves including these data are shown in Fig. \ref{fig:lc}, and the spectral energy distributions (SED) at three epochs are shown in Fig. \ref{fig:sed}.

We have also checked pre-SN ALMA data covering the position of SN 2020oi (proposal IDs: 2013.1.00634S and 2015.1.00978S) \citep{gallagher2018a,gallagher2018b}. We do not detect a point source at the SN location in the continuum emission, with the $1\sigma$ upper limits of 0.078 mJy (100 GHz) and 0.062 mJy (250 GHz). If we consider dust emission at the SN environment, it should follow the Rayleigh-Jeans tail and we may further place an upper limit of $\sim 0.01$ mJy at 100 GHz, which is negligible as compared to the SN emission.

\section{Properties of radio emission from SN 2020oi}\label{sec:lc}

\subsection{Characteristic properties of the synchrotron emission from SNe}\label{subsec:lc1}

In order to analyze the multi-band radio data of SN 2020oi, we summarize basic properties of the synchrotron emission from SNe in this section. Throughout this section, we focus on the case where the CSM density distribution is spherically symmetric and follows a single power law ($\rho_{\rm CSM} \propto r^{-s}$), i.e., the standard assumptions widely adopted in analyzing the radio data of SNe. We are especially interested in clarifying the prediction for the CSM created by a steady-state wind, i.e., $s=2$. In addition to the CSM distribution, below we use a specific (and typical) ejecta structure to show some specific values in the expected radio properties, but we emphasize that the values here are not the primary interest; the key issue here is how the expected properties change as a function of time (e.g., the light curve becoming steeper or flatter), and this behavior is independent from the specific details. 

The synchrotron emission originating in the SN-CSM interaction is characterized by the spectral index ($\alpha$) and temporal slope ($\beta$) ($L_\nu \propto \nu^\alpha t^\beta$). In the adiabatic regime, $\alpha = (1-p)/2$ and $\beta = (3m-3) + (1-p)/2$, where $p$ is the spectral index in the relativistic electron energy distribution. Here, $m$ expresses the evolution of the shock wave as $R_{\rm SN} \propto t^m$ \citep{fransson1998,bjornsson2004,chevalier2006,maeda2012,maeda2013a}, where $R_{\rm SN}$ is the radius at the shock front. For $p = 3$ typically found for SESNe, $\alpha = -1$.  
The self-similar decelerated shock solution predicts $m = (n-3)/(n-s)$, where $n$ is the power-law index in the density distribution within the outer SN ejecta ($\rho_{\rm SN} \propto v^{-n}$) and $n \sim 10$ is frequently adopted for SESNe \citep{chevalier1982,chevalier2006}. Substituting $p=3$ and $m=0.875$ (for $s=2$) into the above equation, we obtain $\beta = -1.375$. 

In the IC cooling regime, adopting $p=3$, the predicted behaviors are $\alpha = -1.5$ and $\beta = (5m-5) - 1/2 - \delta$ \citep{maeda2013a} (after correcting a typo in the reference), where $\delta$ approximates the evolution of the bolometric luminosity (i.e., seed photons) as $L_{\rm bol} \propto t^\delta$. If we adopt $m=0.875$ (to take into account the deceleration) and constant bolometric luminosity (i.e., at the bolometric peak), the expected temporal slope is $\beta = -1.125$. If we instead adopt $m = 1$ (free expansion), then $\beta = -0.5$. Before the peak, $\delta > 0$ and thus the temporal slope is expected to be steeper than the above prediction (due to the increasing number of seed photons). The opposite is true after the bolometric peak ($\delta < 0$). 

\subsection{Analysis of the radio data of SN 2020oi}\label{subsec:lc2}

\begin{figure}[t]
\centering
\includegraphics[width=\columnwidth]{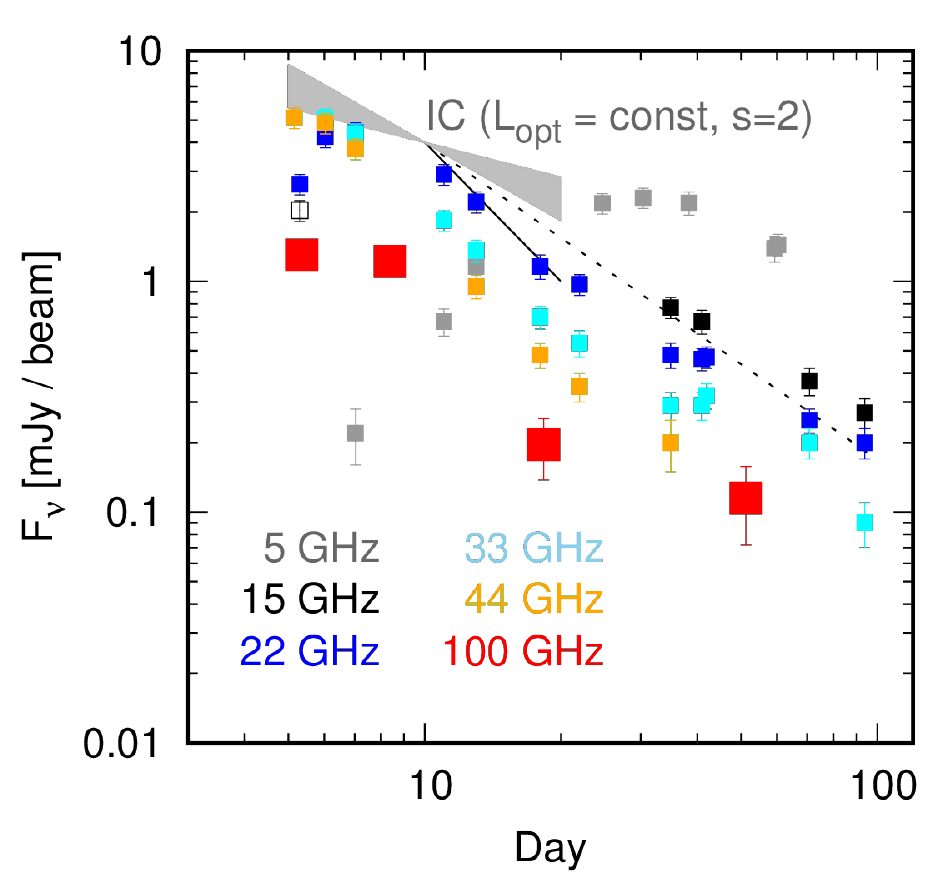}
\caption{The radio light curves of SN 2020oi. The lower-frequency data are from \citet{horesh2020}. The first point for the 15 GHz data (open square) is in fact measured at 16.7 GHz, and is likely an overestimate of the flux at 15 GHz. The flux densities are shown with $1\sigma$ error. A few power-law lines (Section \ref{subsec:lc1}) are shown which intersect at 10 days \citep[roughly the bolometric maximum date]{horesh2020,rho2021}; the dotted black line shows an example of the theoretically expected slope in the adiabatic regime ($\beta=-1.375$), and the gray region represents a range of the expected slope in the IC cooling regime with a constant bolometric luminosity ($\beta = -1.125 \sim -0.5$) \citep{maeda2013a}, adopting the same ejecta and CSM structures with the ones for the adiabatic regime. The black solid line is for $\beta = -2.0$ roughly fitting the slope in the phase immediately after the bolometric maximum. 
}
\label{fig:lc}
\end{figure}

\begin{figure}[t]
\centering
\includegraphics[width=\columnwidth]{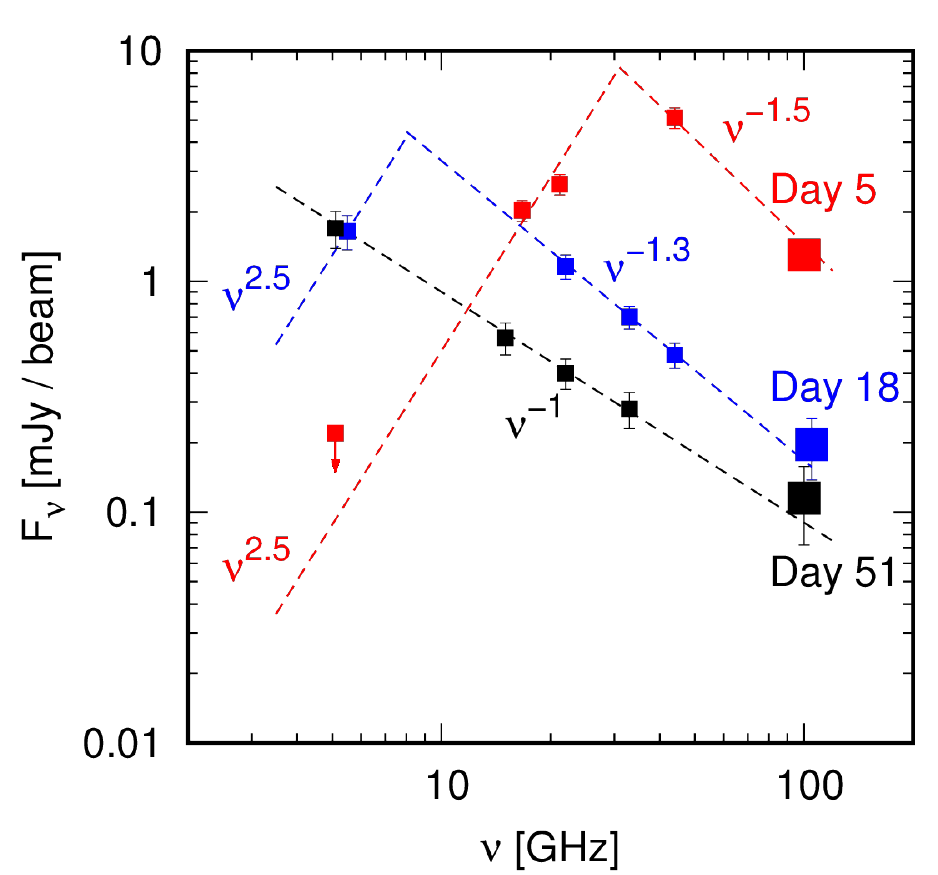}
\caption{The spectral energy distributions at three epochs. The lower-frequency data are from \citet{horesh2020}. The flux densities are shown with $1\sigma$ error. The expected spectral slopes are shown for the optically thick regime ($F_\nu \propto \nu^{2.5}$) and in the optically thin regime ($\propto \nu^{-1}$ to $\nu^{-1.5}$).
}
\label{fig:sed}
\end{figure}

Figure \ref{fig:lc} presents the multi-band radio light curves of SN 2020oi. The ALMA light curve after $\sim 10$ days follows the behavior similar to the light curves in the other wavelengths at $\gsim 15$ GHz but with different flux density level, indicating that SN 2020oi is in the optically thin limit after $\sim 10$ days except for the lowest frequency. The ALMA data are unique at $\lsim 10$ days; the spectral energy distribution (Fig. \ref{fig:sed}) shows that SN 2020oi is fully optically thin at 100 GHz even at $\lsim 10$ days, while it is not the case in the lower frequencies (e.g., it is only marginally optically thin at 44 GHz at day 5). This characteristic nature at the high frequency provides a powerful tool to investigate the nature of the CSM; the time evolution here should directly reflect the CSM density structure as summarized below (but see Section 5 for caveats and other possibilities).

The radio evolution of SN 2020oi in the optically-thin wavelengths (i.e., at 100 GHz for $\lsim 10$ days and at $\gsim 15$ GHz for $\gsim 10$ days) can be divided into three characteristic phases (Fig. 2); a flat evolution in the early phase ($\lsim 10$ days), a rapid decay in the intermediate phase ($\sim 10-40$ days), and a slow decay in the late phase ($\gsim 40$ days). 

As described in Section 3.1, the spectral index of $\alpha \sim -1$ ($f_\nu \propto \nu^\alpha t^\beta$) seen in the late phase is the one predicted for the adiabatic case with the relativistic electron energy distribution having the power-law index of $p \sim 3$ \citep{bjornsson2004,chevalier2006,maeda2013a}. For the CSM density distribution described as $\rho_{\rm CSM} \propto r^{-s}$ and $s=2$ (i.e., steady-state mass loss), the theoretically predicted power-law index in the light curve decline, in any optically-thin band, is $\beta = -1.375$ for the SN ejecta outer layer having a typical distribution of $\rho_{\rm SN} (v) \propto v^{-10}$ where $v$ represents the ejecta velocity coordinate (see Section 3.1). This slope matches to the observed light curve evolution in the late phase ($\gsim 40$ days) reasonably well without fine tuning. These properties in the late phase are typical for radio emission from SESNe at similar phases \citep{chevalier2006}.

In the earlier phases ($\lsim 40$ days), the optically thin SED passing through the ALMA band is much softer, reaching to $\alpha \sim -1.5$. This spectral slope derived from the multi-band data is consistent with the SED within the ALMA band 3, where we see a clear trend of decreasing flux toward the higher frequency; however the individual spectral window data are not very useful to constrain the spectral slope given the limited frequency range, and we rely on the slope derived through the multi-band data\footnote{The analysis of the individual spectral windows for the third and fourth epochs is even less constraining; the expected flux change within the band is $\sim 10$\%, which is already below the 1$\sigma$ error in the data obtained by combining all the spectral windows (Tab. \ref{tab;flux}).}. The spectral slope is fully consistent with the spectral steepening due to the electron cooling, very likely caused by the inverse Compton (IC) cooling \citep{horesh2020} (see Section 5 for further details). The flux at $\lsim 40$ days is suppressed as compared to a simple extrapolation from the later phase, as expected by the increasing importance of the IC cooling effect toward the earlier epochs. 

However, a detailed investigation raises a complication. In Figure 2, the expected range of the slope for the light curve in the IC-dominating regime is shown by a gray region around the bolometric luminosity maximum (i.e., $\sim 10$ days) covering the free expansion and the decelerated shock cases, for the CSM with $s=2$ and a constant bolometric luminosity for the IC seed photons \citep{maeda2013a} (see also Section 3.1). The expected slope should be steeper/flatter than this slope before/after the bolometric maximum, due to the increasing/decreasing number of the seed photons. 

The observed behavior is opposite to this prediction. In the early phase before the optical peak ($\lsim 10$ days), the optically thin ALMA data show a flatter evolution than predicted. In the intermediate phase after the optical peak ($\sim 10 - 40$ days), the observed multi-band light curves ($\gsim 22$ GHz for the optically thin emission) are steeper. It strongly suggests that the CSM structure in the vicinity of SN 2020oi deviates substantially from a single power-law distribution; a flat distribution in the innermost region and the usual steady-state distribution in the outermost region are connected by a steep density decrease in between. 

While the prediction above adopts a specific combination for the CSM density distribution ($s=2$) and the ejecta density structure ($n=10$) as motivated by the late-phase behavior, we emphasize that the detail here is not important; for example, if one would adopt a combination leading to the steeper light curve evolution (even if such evolution would not explain the late epoch), one could fit to the optical post-peak decline but the discrepancy in the pre-peak becomes more significant. The opposite is true if one would adopt the combination leading to a flat evolution in the intrinsic light curve. The key point is that the discrepancy from the predicted behavior is in the opposite direction before and after the optical peak, which is not remedied by changing the CSM and ejecta density distribution as long as the single power-law function is adopted. 

\section{Light Curve Models and the CSM environment}\label{sec:model}

We further quantitatively investigate the CSM distribution with two-steps light curve model calculations. At the first step (Section \ref{subsec:model1}), the multi-band synchrotron light curves are computed for a `single power-law' CSM density structure, and the model is applied to each segment of the light curve evolution (i.e., the early, intermediate, and late phases) separately. This exercise confirms the need for the non-smooth CSM distribution. The (approximately) derived CSM structure is then used to directly compute the multi-band light curves for an arbitrary (but spherically symmetric) CSM structure in the second step (Section \ref{subsec:model2}). We have further modified the input CSM structure for the improvement of the model light curves as compared to the data. 

The distance to M100 has been derived to be $\sim 14-20$ Mpc by various measurements\footnote{https://ned.ipac.caltech.edu.}, and we adopt 15.5 Mpc. 
The distance adopted in the previous works for SN 2020oi is in the range of $\sim 14-17$ Mpc, i.e., $15.5 \pm 1.5$ Mpc. We may thus consider that the uncertainty in the distance is $\pm 10$\%. Combining rough constraints (i.e., scaling relations) placed on the synchrotron self-absorption frequency, IC cooling frequency and the optically thin synchrotron emission flux \citep[e.g., ][]{maeda2012}, we estimate that the distance uncertainty translates to the uncertainty in deriving values of the microphysics parameters ($\epsilon_{\rm e}$ and $\epsilon_{\rm B}$; see Section 4) and the CSM density scale (i.e., the mass-loss rate) at the level of a factor of two.  This is sufficiently small for the purpose of the present work. Further, we emphasize that the analysis of the time evolution, which is the main focus of the present work, is essentially independent from the distance uncertainty. 

\begin{figure*}[t]
\centering
\includegraphics[width=0.65\columnwidth]{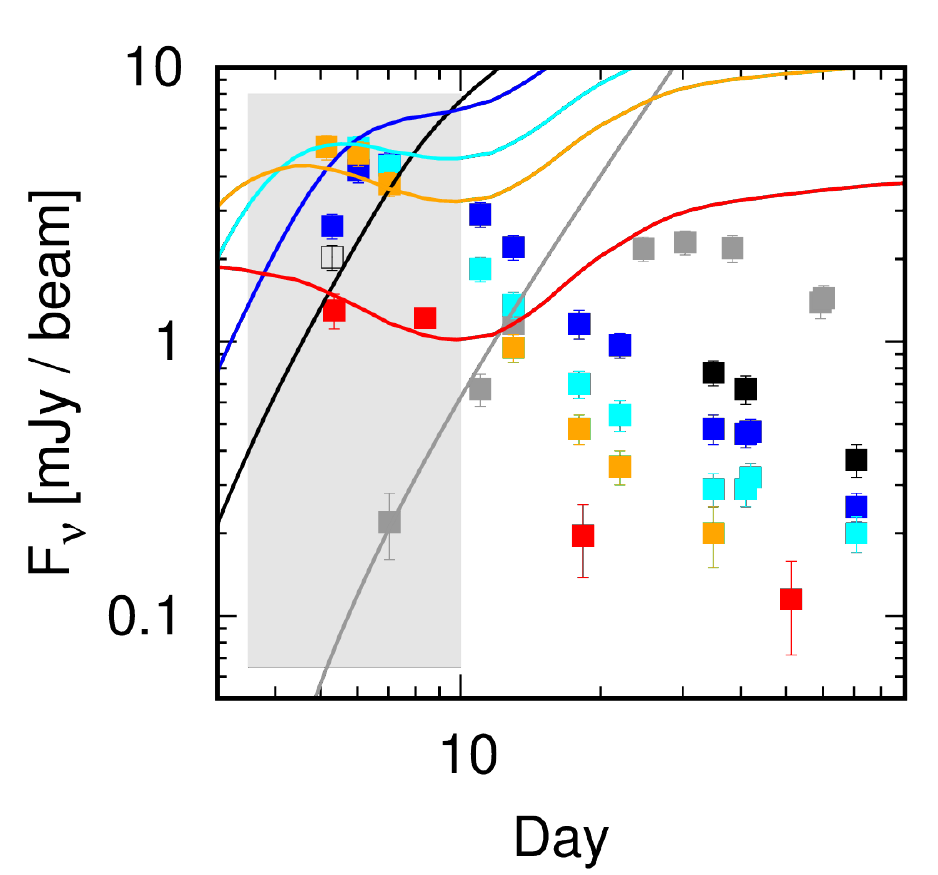}
\includegraphics[width=0.65\columnwidth]{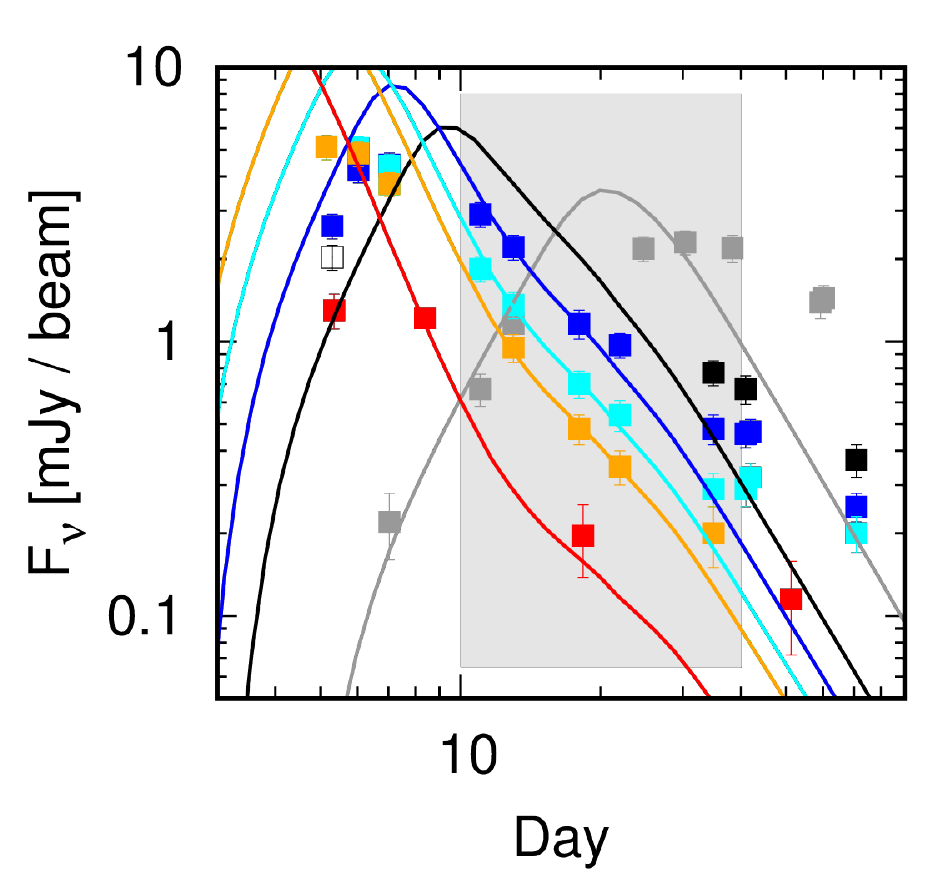}
\includegraphics[width=0.65\columnwidth]{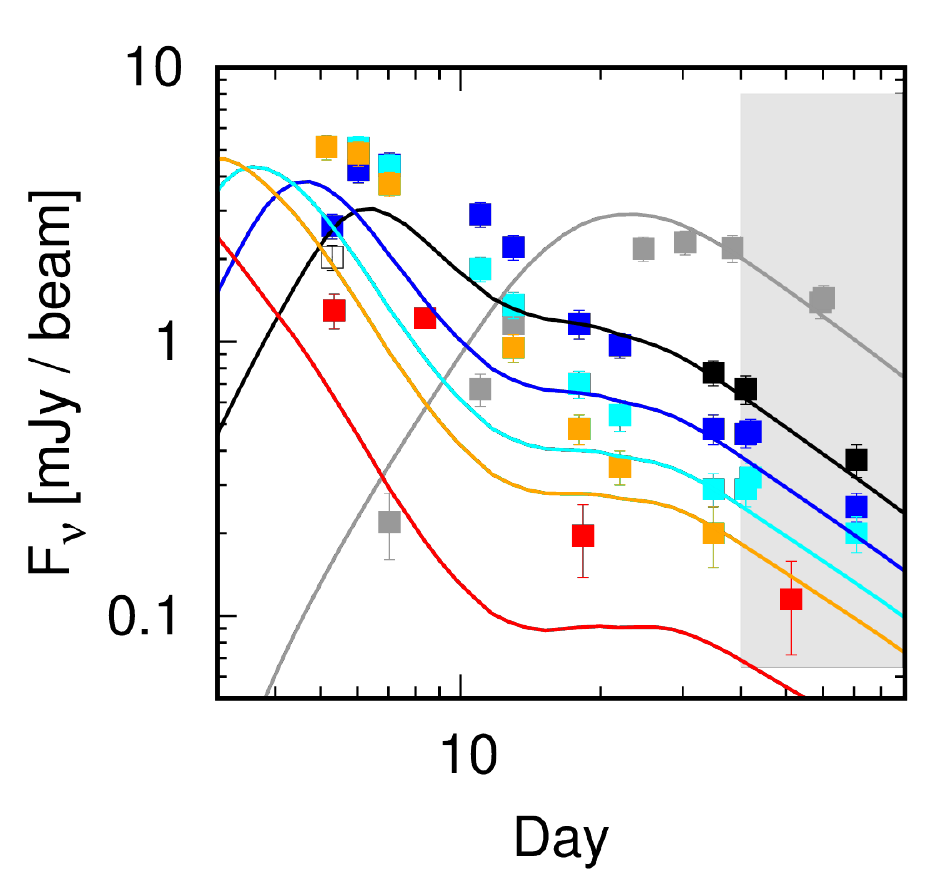}
\caption{The first-step model light curves as compared to the multi-band light curves of SN 2020oi. The three models are shown, for the early phase with $\rho_{\rm csm} \propto r^{-1.5}$ (left), for the intermediate phase with $\propto r^{-3}$ (middle), and for the late phase with $\propto r^{-2}$ (right), each of which applies only to the limited time window (shaded area in each panel). The color scheme for the data symbols is the same as for Fig. \ref{fig:lc}, and the model curve at each frequency has the same color as the corresponding data symbol.
}
\label{fig:toymodel}
\end{figure*}

\begin{figure}[t]
\centering
\includegraphics[width=\columnwidth]{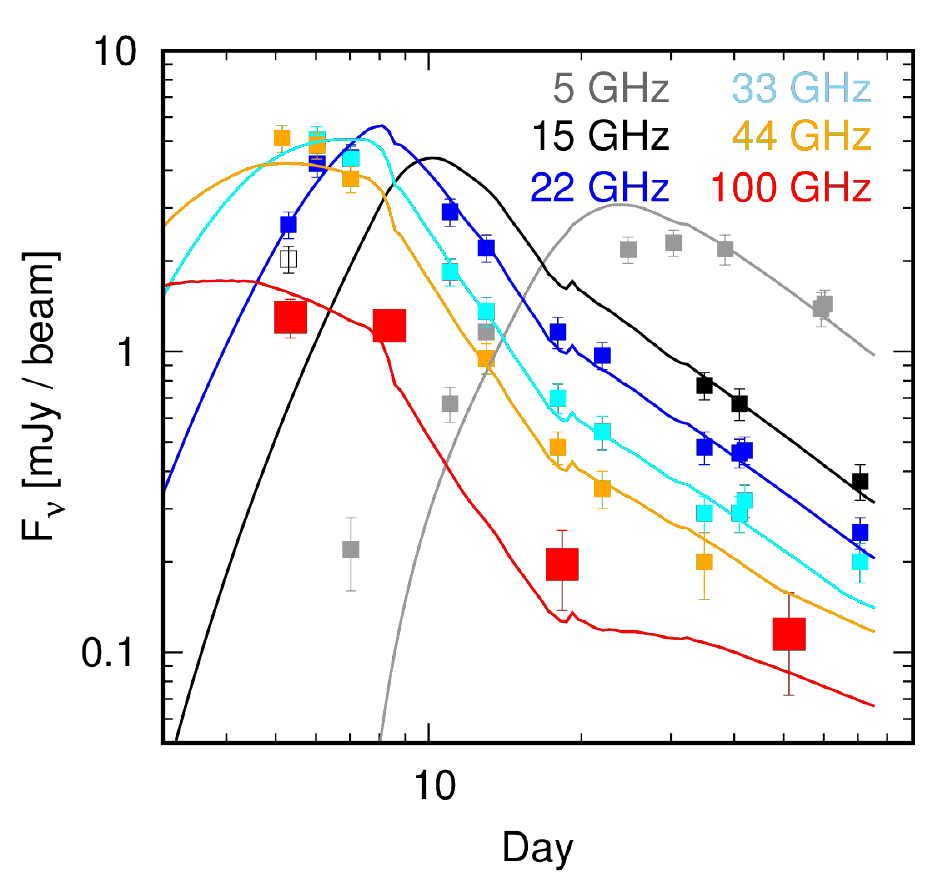}
\caption{Model light curves as compared to the multi-band light curves of SN 2020oi. The model here is computed with the hydrodynamic simulation with the CSM density structure given in Fig. \ref{fig:csm}. The color scheme for the model curves is the same as that for the data symbols indicated by the labels.
}
\label{fig:model}
\end{figure}

The results from the first-step and the second-step models are shown in Figs. \ref{fig:toymodel} and \ref{fig:model}, respectively. The CSM density distribution thus inferred is shown in Fig. \ref{fig:csm}. We emphasize that these light curve models are performed to confirm the need for the non-smooth CSM, and to demonstrate that the CSM structure, qualitatively inferred through the analyses of the key physical processes involved in the synchrotron emission (Section \ref{sec:lc}), does reproduce the characteristic nature of SN 2020oi seen in the radio data, rather than aiming at deriving the CSM structure accurately. 

The initial conditions for both models are the structures of the SN ejecta and the CSM. Adopting a broken power law for the ejecta density structure, the properties of the SN ejecta are specified by the ejecta mass ($M_{\rm ej}$), kinetic energy ($E_{\rm K}$), and the power-law indices of the inner and outer density distributions in the ejecta velocity coordinate ($v$); the inner index is set to be 0, and the outer one is denoted as $n$ ($\rho_{\rm SN} \propto v^{-n}$). We adopt $M_{\rm ej} = 1 M_\odot$, $E_{\rm K} = 10^{51}$ ergs, and $n=10$, being roughly consistent with the optical light curve modeling \citep{rho2021}. We emphasize that the details of these parameters on the ejecta properties are not important for our purpose, which is to demonstrate the need for the non-smooth CSM as argued in Section \ref{sec:lc} in a model-independent way. Changing the ejecta properties could change the overall flux level of the synchrotron emission and thus would affect the overall density scale of the CSM derived by the comparison between the model and observation; however, it is the temporal `evolution' that matters for the present propose.

Once the properties at the shock front ($R_{\rm SN}$ for the radius and $V_{\rm SN}$ for the velocity) are obtained, we use the standard formalism widely used for simulating the synchrotron emission from the SN-CSM interaction \citep{fransson1998,bjornsson2004,chevalier2006,maeda2012,matsuoka2019}, under the widely-used assumptions that the accelerated electrons have a power-law energy distribution (with the index of $p$), and that certain fractions of the energy, $\epsilon_{\rm e}$ and $\epsilon_{\rm B}$, dissipated at the shock are  transferred to the energy of the relativistic electrons and the amplified magnetic field, respectively. For the cooling processes, we take into account both the synchrotron and inverse Compton (IC) cooling. For the latter, we adopt the bolometric light curve presented by \citet{horesh2020}. The synchrotron self absorption (SSA) is taken into account \citep{chevalier1998}. The free-free absorption (FFA) is not important in the present study, but for completeness it is included for the He-rich composition \citep{matsuoka2019}. For the FFA, the pre-shock CSM temperature is uncertain, and we take $\sim 10^{5}$ K. The microphysics parameters are $p$, $\epsilon_{\rm e}$, and $\epsilon_{\rm B}$.

\subsection{A single-power law CSM distribution for each epoch}\label{subsec:model1}

In the first-step model, we adopt a single power law in the form of $\rho_{\rm csm} = D r^{-s}$ for the CSM density distribution. The parameters here, $D$ and $s$, are the main targets to derive/estimate through the radio emission modeling. For the first two phases (up to 40 days), we assume a constant velocity for the shock wave, $V_{\rm SN} = 30,000$ km s$^{-1}$ \citep{horesh2020}, since the swept-up mass is not sufficient to decelerate the shock wave \citep{maeda2013a}. For the late epoch, the self-similar solution describing the deceleration of the shock wave is used \citep{chevalier1982}. We note that the velocity evolution here is adopted for a demonstration purpose; this will be numerically solved in the second-step model. 

The situation becomes progressively simpler toward the later epoch, where the effects of absorption and cooling become negligible. We thus start with the model for the late epoch, and then move to the earlier epochs. The model light curves are shown in Figure \ref{fig:toymodel}, and the density structure used in the model is shown in Fig. \ref{fig:csm}. 

The right panel of Figure \ref{fig:toymodel} shows the model applied to the late phase. We adopt $s=2$ for the CSM structure. For the microphysics parameters, we adopt $p=3$, $\epsilon_{\rm e} = 0.01$, and $\epsilon_{\rm B} = 0.005$ (see also Section \ref{sec:discussion}). These parameters are consistent with those obtained for SN IIb 2011dh by a combined analysis of radio and X-ray data \citep{maeda2012,maeda2014}. The same model extended to the early phase (non-shaded region in the same panel of Fig. 4) however illustrates the problems mentioned in Section \ref{sec:lc}. It is clear that the IC cooling is at work, but the temporal evolution is not recovered with $s = 2$.

The need for the steeper density gradient for the intermediate phase is thus clear. The middle panel of Fig. \ref{fig:toymodel} shows our model for this phase, with $s = 3$ instead of $s = 2$. The model explains the data reasonably well for $\sim 10 - 40$ days. This model, again, has a problem if this would be further extended down to the earlier phase ($\lsim 10$ days). For example, the ALMA data readily reject the applicability of the same CSM structure before $\sim 10$ days; it should already be in the optically thin regime (Fig. \ref{fig:sed}), and the steep density distribution results in too rapid decay.  

Therefore, we have introduced a flat density structure in the innermost CSM. The left panel of Fig. \ref{fig:toymodel} shows the model with $s = 1.5$. The model can explain the increasing trend in the lower frequency bands and the flat evolution in the higher frequency bands, and the spectral index is largely consistent with the data.

The microphysics parameters are set identical among the models for the different phases. The electron energy distribution, $p \sim 3$, is robust; the combination of this with the simplest CSM with $s = 2$ naturally explains the late-phase light curves. This is typical for SESNe in the similar phases \citep{chevalier2006}. We then have basically three parameters to characterize the radio emission; $\epsilon_{\rm e}$, $\epsilon_{\rm B}$, and the CSM density. We have at least three independent observational constraints; effect of the SSA (in the early phase), effect of the IC cooling (in the intermediate phase), and the optically thin flux (in the late phase).  Therefore, the model parameters are not seriously degenerate. In any case, the exact values of $\epsilon_{\rm e}$ and $\epsilon_{\rm B}$ are not the main focus in the present paper; our main interest in the present work is the temporal evolution of the radio properties, which is then translated to the CSM radial distribution. It is largely insensitive to the values of these parameters (as long as they are constant in time; see Section \ref{sec:discussion}).

\subsection{A detailed modeling with the non-smooth CSM distribution}\label{subsec:model2}

As the second step, we have performed a more detailed analysis, by numerically solving the evolution of the shock wave for a non-smooth CSM density distribution. The hydrodynamic evolution is solved using SNEC (the SuperNova Explosion Code) \citep{morozova2015}\footnote{https://www.stellarcollapse.org/SNEC.}. The simulation provides the shock evolution, i.e., $R_{\rm SN}$ and $V_{\rm SN}$. The information is then used to compute the synchrotron emission. The SN ejecta properties and the microphysics parameters are set the same with the first-step model. An exception is the FFA, for which we changed the pre-shock temperature to see if improvement can be obtained, and the final value we adopt is $2 \times 10^{5}$ K. As another modification, we have introduced a flattening of the electron energy distribution to $p = 2.1$ above the Lorentz factor $300$, where the electrons have sufficiently high energy to be further accelerated efficiently \citep{maeda2013b}; this slightly enhances the late-time high-frequency emission and provides a better agreement to the data.

The CSM density distribution obtained in the first step is used as an initial guess in the second step. The two models do not necessarily agree; the hydrodynamic evolution adopted in the first step is much less accurate, given the deviation of the CSM structure from a single power-law distribution. Therefore, we have further tuned the input CSM structure. The result of this exercise is shown in Figs. \ref{fig:model} and \ref{fig:csm}. This detailed model confirms the robustness of the physical interpretation and constraints provided in a model-independent way (Section \ref{sec:lc}) and by the first-step model (Section \ref{subsec:model1}), and explain the multi-band light curves of SN Ic 2020oi reasonably well.

\section{Discussion}\label{sec:discussion}

\subsection{The mass-loss history and the implications for the pre-SN activity}\label{subsec:massloss}

\begin{figure}[htbp]
\centering
\includegraphics[width=\columnwidth]{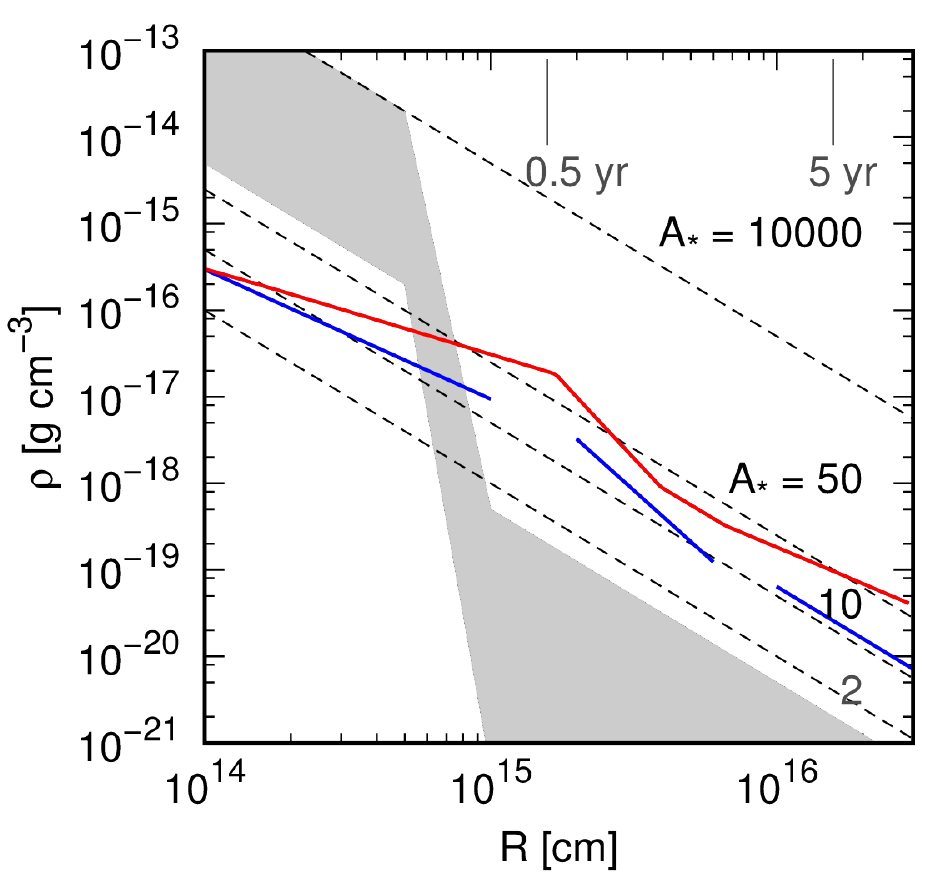}
\caption{The circumstellar density distribution inferred for SN Ic 2020oi. The red line is for the final light-curve model (Fig. \ref{fig:model}), while the blue lines are for the first-step model applied to each time window (Fig. \ref{fig:toymodel}). For comparison, the fiducial CSM structure derived for SN II 2013fs is shown by the gray shaded area \citep{yaron2017}. The CSM distributions for steady-state mass loss with a constant velocity (i.e., $\rho_{\rm csm} = 5 \times 10^{11} A_{*} r^{-2}$, where $A_{*} = (\dot M / 10^{-5} M_\odot {\rm yr}^{-1}) / (v_{\rm w} / 1,000 {\rm \ km \ s}^{-1})$) are shown for different values of $A_{*}$. The corresponding look-back time in the mass-loss history before the explosion is indicated for 0.5 and 5 yr for $v_{\rm w} = 1,000$ km s$^{-1}$. 
}
\label{fig:csm}
\end{figure}

\begin{figure}[t]
\centering
\includegraphics[width=\columnwidth]{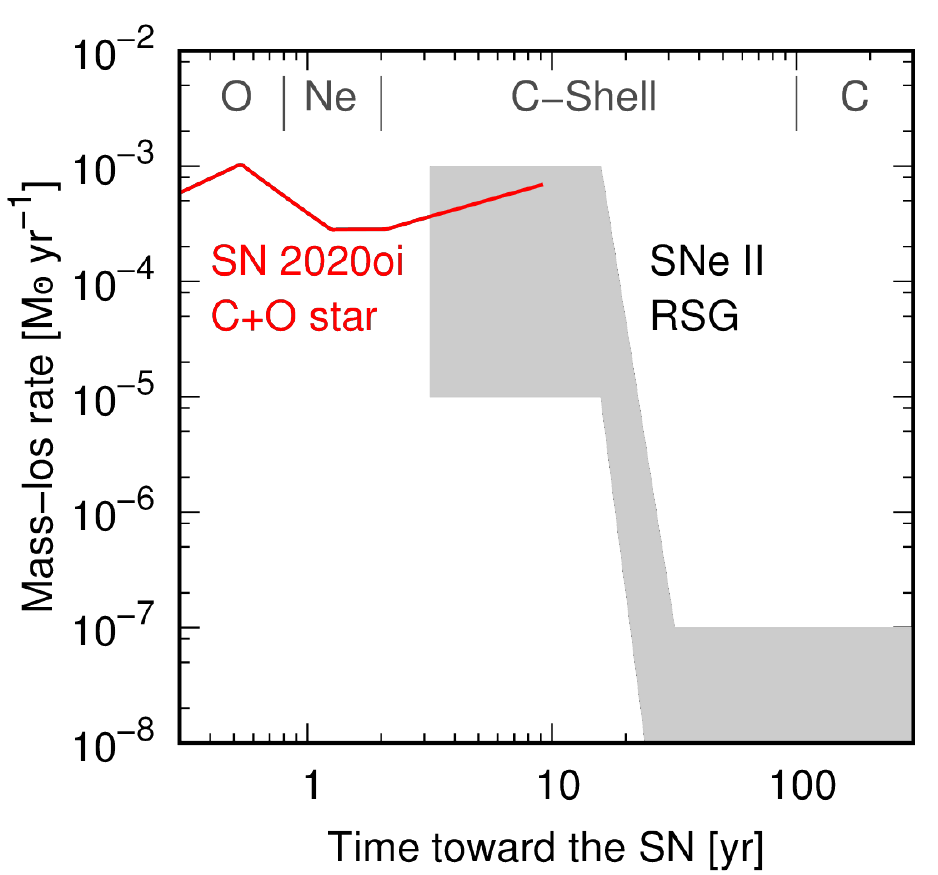}
\caption{The mass-loss history inferred for SN Ic 2020oi. The CSM density distributions are converted to the mass-loss history, assuming $v_{\rm w} = 1,000$ km s$^{-1}$ for SN Ic 2020oi (red line) and $10$ km s$^{-1}$ for SN II 2013fs (gray shaded area). The change in the pre-SN nuclear burning stage is indicated on the top for the star with the initial main-sequence mass of $\sim 15 M_\odot$ \citep{fuller2017}. 
}
\label{fig:mdot}
\end{figure}

Fig. \ref{fig:csm} shows the comparison of the CSM structures derived for SN Ic 2020oi in this work and for a representative SN II inferred from the optical data \citep{yaron2017}. The present work probes the CSM distribution at $\sim 10^{15}$ cm as is similar to the previous works for SNe II. However, the corresponding look-back time in the mass-loss history is different; we are able to reach the look-back time of $\lsim 1$ yr for SN Ic 2020oi. In Fig. \ref{fig:mdot}, the CSM density as a function of radius is converted to the mass-loss rate as a function of the pre-SN look-back time assuming a constant mass-loss wind velocity (adopting $v_{\rm w} = 1,000$ km s$^{-1}$ for SN Ic 2020oi and $10$ km s$^{-1}$ for SN II 2013fs).

The timing in the changes in the mass-loss property derived here roughly corresponds to the transitions in the nuclear burning stages; from the carbon to the neon burning and then the neon to the oxygen burning \citep{langer2012,fuller2017}. The change in the pre-SN nuclear burning stage is indicated in Fig. \ref{fig:mdot} for the star with the initial main-sequence mass of $\sim 15 M_\odot$ \citep{fuller2017}. The main-sequence mass for SN 2020oi is suggested to be $\sim 13 M_\odot$ by \citet{rho2021} through the optical light curve model, and the time scale for the advanced burning stages is similar. A somewhat smaller progenitor mass ($\sim 9.5 M_\odot$) has been suggested by \citet{gag2021} through a similar approach; the time scale for the advanced burning stages is then longer by a factor of about two than shown in Fig. \ref{fig:mdot} \citep{jones2013}. As an extreme case, the time scale for the advanced burning stages is shorter by a factor of a few for a $25 M_\odot$ star than a $15 M_\odot$ star \citep{limonji2000}.  In summary, irrespective of the progenitor mass, the expected time scale for the advanced burning stages fits into the time scale we have derived.
The `confined CSM' derived for SNe II may also correspond to the transition from the carbon core burning to the shell burning. 

These findings suggest that the change in the mass-loss properties in the final evolution of massive stars is likely driven by the change in the nuclear burning stage. Since the time scale here is much shorter than the thermal time scale of a progenitor C+O star ($\sim 1,000$ yrs), the whole star should respond dynamically to the change in the nuclear burning stage in the core \citep{ouchi2019,morozova2020}. This thus raises a need for new generation stellar evolution theory to fully understand the final evolution, beyond the classical quasi-static theory.

We note that it is also possible that the change in the mass-loss behavior is driven by the change in the mass-loss velocity rather than the mass-loss rate. This would not change our main conclusion, as it should also indicate that the characteristic time scale for the pre-SN activity is at most one year, overlapping with the nuclear burning time scale at the end of the stellar life. 

Comparison between SESNe and SNe II in the derived mass-loss history is intriguing, with a caveat that the difference might also be contributed by the difference in the envelope structure. The derived mass-loss rate for SN Ic 2020oi, assuming $v_{\rm w} \sim 1,000$ km s$^{-1}$, is $10^{-4} - 10^{-3} M_\odot$ yr$^{-1}$. This is high as compared to the standard steady-state wind for a C+O star \citep[$\sim 10^{-5} M_\odot$ yr$^{-1}$;][]{crowther2007}, but the fluctuation in the mass-loss rate in the final few years is indeed modest and less than an order of magnitude. The mass-loss rate in this final burning stage is roughly at the same order to the mass-loss rate derived for SNe II in the less advanced stage at $\sim 10$ yrs before the explosion (Fig. \ref{fig:mdot}). 

\subsection{The effect of the Inverse Compton (IC) cooling and constraints on the microphysics parameters}\label{subsec:ic}

In this section, we provide a qualitative estimate on the effect of the IC cooling. This is used to check the consistency of the microphysics parameters adopted in Section \ref{sec:model}. In the following, the values mentioned for some physical quantities are used only for an order-of-magnitude estimate, frequently adopted from the `results' from the light curve modeling (Section \ref{sec:model}). Note that these values are not `assumed' in the light curve models in Section \ref{sec:model}. 

\begin{figure}[t]
\centering
\includegraphics[width=\columnwidth]{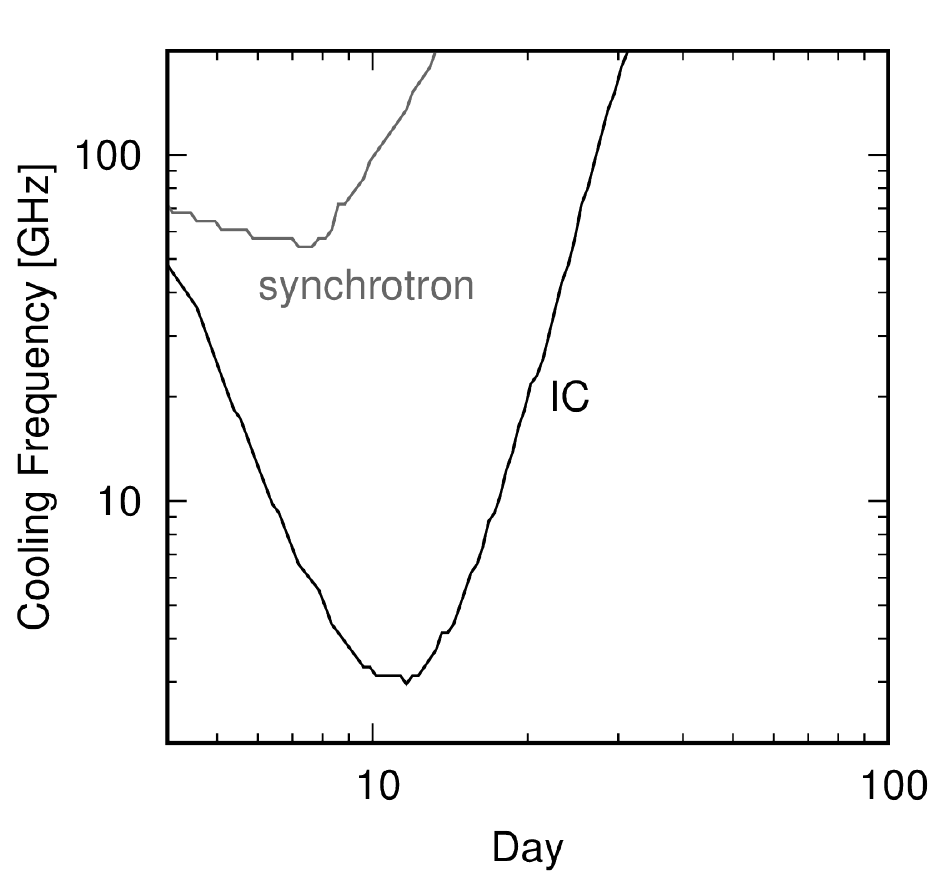}
\caption{Effect of the IC cooling. The cooling frequency (above which the synchrotron SED becomes steeper) is shown for the IC (black) and for the synchrotron cooling (gray). The figure is for our `final' model.
}
\label{fig:ic}
\end{figure}

The ratio of the synchrotron cooling time scale ($t_{\rm syn}$) and the IC cooling time scale ($t_{\rm IC}$) is roughly expressed as follows: $t_{\rm syn} / t_{\rm IC} \sim 65 B^{-2} L_{{\rm bol}, 42} R_{15}^{-2}$, where $B$ is the amplified magnetic field strength in Gauss, $L_{{\rm bol}, 42}$ is the bolometric luminosity in $10^{42}$ erg s$^{-1}$, and $R_{15} = R_{\rm SN} / 10^{15}$ cm \citep{bjornsson2004,maeda2013a}. At the bolometric peak ($\sim 10$ days), $L_{{\rm bol},42 } \sim 2$ \citep{horesh2020,rho2021} and we may take $R_{15} \sim 2$ for $V_{\rm SN} \sim 30,000$ km s$^{-1}$. Therefore, the IC cooling becomes a dominant cooling process if $B \lsim 6$ Gauss. Given $B = \sqrt{8 \pi \epsilon_{\rm B} \rho_{\rm csm} V^2}$ with $V \sim 30,000$ km s$^{-1}$, the condition, $\epsilon_{\rm B} \lsim 0.016 (\rho_{{\rm csm}, -17})^{-1}$, must be satisfied for the IC cooling to dominate over the synchrotron cooling. Here $\rho_{{\rm csm}, -17}$ is the CSM density at $R_{15} \sim 1$ as normalized by $10^{-17}$ g cm$^{-3}$ as found in our final result. This is consistent with $\epsilon_{\rm B} = 0.005$ adopted in our models. 

Indeed, in our model (Fig. \ref{fig:model}) we find $B \sim 2$ Gauss at 10 days. By adopting $B\sim 2$ Gauss and the other characteristic values for SN 2020oi, we obtain the expected luminosity at 100 GHz in the optically thin, IC cooling regime, as $\sim 4 \times 10^{23} n_{\rm e}$ erg s$^{-1}$ Hz$^{-1}$, where $n_{\rm e}$ is the number density of the relativistic electrons and $n_{\rm e} \sim \epsilon_{\rm e} \rho_{\rm CSM} V_{\rm SN}^2/m_{\rm e} {\rm c}^{2} \sim 600 (\epsilon_{\rm e}/0.01)$ cm$^{-3}$ \citep{bjornsson2004,maeda2013a}. The observed synchrotron flux at $\sim 10$ days corresponds to $\sim 2 \times 10^{26}$ erg s$^{-1}$ Hz$^{-1}$ at 100 GHz. Therefore, $\epsilon_{\rm e} \sim 0.01$ is required.

The estimate here shows $\epsilon_{\rm e} \sim \epsilon_{\rm B} \sim 0.001 - 0.01$, being consistent with the values adopted in our model. Fig. \ref{fig:ic} shows the evolution of the cooling frequencies found in our final model (Section \ref{subsec:model2}; Fig. \ref{fig:model}), showing that the IC cooling does dominate in the model. The deviation from the full equipartition is preferred, confirming the earlier claim \citep{horesh2020}. We however do not require an extremely large deviation as suggested by \citet{horesh2020}, which might highlight the importance of the multi-wavelength and time-dependent model calculation, including various cooling and absorption processes simultaneously, as we conducted in the present work. In any case, we emphasize that the ratio of $\epsilon_{\rm e}$ and $\epsilon_{\rm B}$ can change the overall CSM density scale but not the normalized CSM density structure as a function of radius, and thus would not affect our main conclusion on the need for the non-smooth CSM structure. 

\subsection{Other possible explanations}\label{subsec:caveats}

In the present work, we have shown that the evolution of the multi-band light curves of SN Ic 2020oi, starting at the infant phase, is not explained by CSM distributed smoothly in the radial direction. This conclusion has been reached based on the light curve analysis and modeling under the standard assumptions widely adopted for modeling radio emission from SNe. Especially important assumptions are (1) constant efficiency of the electron acceleration and magnetic field amplification as a function of time, and (2) spherically symmetric CSM distribution. 

The treatment of the microphysics parameters is phenomenological (not only in this study but also in the standard analysis for radio SNe), as the details of the particle acceleration are not well understood yet. This is especially the case for the acceleration of electrons to the relativistic energy, known as the electron injection problem \citep[see][for a particular case of an SN-induced shock]{maeda2013b}. However, previous works have been largely successful to explain the behaviors of synchrotron emission from various types of SNe including SESNe, under the prescription similar to (or the same with) the present work assuming the constant efficiency for the electron acceleration and magnetic field amplification \citep[e.g.,][]{chevalier2006}. Also, the validity of this approximation has been confirmed for a particular case of SN IIb 1993J with a more detailed analysis \citep{fransson1998}. Further, if the possible temporal evolution of the microphysics parameters would explain the light curves of SN 2020oi, it would probably require a complicated non-monotonic evolution as a function of time to explain the characteristic `flat-steep-flat' evolution seen in the radio light curves of SN 2020oi (Section \ref{sec:lc}). These considerations suggest that this is not likely the cause of the characteristic evolution seen in SN 2020oi. However, since the SN radio emission in the infant phase has not been well explored both observationally and theoretically, this would require further study.  

Asymmetry (i.e., asphericity) may indeed be an important factor to characterize the CSM around SN progenitors. This may especially be the case for SESNe, for which the binary interaction is a popular scenario for the progenitor evolution \citep{ouchi2017,yoon2017}, which may create either bipolar-type or disk-like CSM distribution. However, in the standard binary evolution model, the binary mass transfer in the final phase is not important for the compact SESN progenitors including SNe Ic \citep{ouchi2017,yoon2017}, and the final mass loss is expected to be dominated by the stellar wind/activity. In addition, most of the previous works for modeling radio emission from SESNe (or SNe in general) assume spherical symmetry, and they are largely successful to explain the SN radio properties \citep[e.g.,][]{chevalier2006}. As such, it is not likely that the characteristic temporal evolution seen in the radio emission from SN 2020oi is originated in the asymmetric CSM. Furthermore, the SED of SN 2020oi at each epoch can be described by a single component, with a broken power law describing the optically thick and thin regimes. If the characteristic behavior in the multi-band light curves is to be explained by the convolution of multiple CSM components, it will result in a complicated SED at least in a transitional phase. Such an evolution is not seen in the data. In summary, we conclude that possible asymmetry is not likely a main cause to create the characteristic radio behavior seen for SN 2020oi. However, the radio emission based on an asymmetric CSM has been largely unexplored, and indeed the effect of asymmetric CSM distribution is an interesting topic for various classes of SNe. We postpone further investigation to the future. 

As yet another caveat, one may ask what if the synchrotron cooling would indeed dominate the cooling process. First of all, as shown in Section \ref{subsec:ic} and by \citet{horesh2020}, the IC cooling is very likely the dominant process. This explains the radio light curve and the SED evolution of SN 2020oi under reasonable physical properties (e.g., in the typical CSM density scale). In addition to these arguments, we also emphasize that it would not help explain the characteristic evolution of SN 2020oi anyway; if the smooth CSM distribution would be assumed, the synchrotron cooling frequency would evolve just monotonically as a function of time. It would then never explain the `flat-steep-flat' evolution of SN 2020oi in its multi-band light curves. Therefore, we would reach the same conclusion, i.e., the need for the non-smooth CSM. One may further consider a very fine-tuned situation where the synchrotron cooling would dominate in the first 10 days and then it would be replaced by the IC cooling. This would remedy the problem in the early-phase flat evolution, but the problem still remains in the post optical-peak behavior (i.e., the light curve in the cooling regime but with the decreasing cooling effect is steeper than the latter adiabatic phase, contrary to the expectation).

\subsection{Importance of the quick radio follow-up observations}\label{subsec:importance}

Studying emission properties from infant SESNe provides a unique opportunity to study the mass-loss history in the final phase before the SN explosion; the CSM at $\sim 10^{15}$ cm that can be probed by the SN emission properties corresponds to the final year (or below one year) for SESNe thanks to the expected high mass-loss wind velocity ($v_{\rm w}$). The present study shows that the quick radio follow-up observation is a powerful method, which can be more efficient to trace the nature of the CSM than the optical observations. 

Based on the CSM density distribution we have derived for SN 2020oi, we estimate the optical depth of the CSM within $\sim 2\times 10^{15}$ cm to be $\sim 0.025 (\kappa/0.2 \ {\rm cm}^2 {\rm g}^{-1}) \ll 1$ for the electron scattering, and thus little trace will be seen at optical wavelengths. It leaves the quick radio follow-up observations, especially at the high frequency to catch the optically thin emission, as a unique tool to probe the mass-loss history down to the sub-year scale before the SN \citep{matsuoka2019}. Given the decreasing synchrotron flux toward the higher frequency, coupled with the crowded environment where core-collapse SNe explode, the ALMA with the combination of high sensitivity and spatial resolution serves as a unique facility. 

There is no signature associated with the CSM seen in the optical/UV data of SN 2020oi at $\gsim 5$ days \citep{rho2021,gag2021}, i.e., the temporal window covered by our radio analysis, supporting the above statement. \citet{gag2021} report a possible enhanced emission in the optical/UV fluxes at $\sim 2.5$ days. The CSM density further extrapolated down to the inner region from that derived by the present study would not account for such emission. If the early optical/UV enhancement would be associated with the SN-CSM interaction, it would require a huge increase in the mass-loss rate in the final few months before the explosion at least by an order of magnitude (see the estimate of the optical depth above), which may be associated with the rapid increase of the nuclear energy generation toward the final Si burning stage. However, the optical/UV data do not allow us to discriminate between different scenarios for the origin of the early emission, and it may not be associated with the CSM at all \citep{gag2021}. This highlights the importance of the rapid radio follow-up observations, especially at high frequency bands; we need very rapid radio follow-up observations of SNe within the first few days to further constrain the final evolution of massive stars. 

\section{Summary}\label{sec:summary}

While there has been accumulating evidence that a large fraction of massive stars experience dynamic activity toward the end of their lives, the investigation has been largely limited to SNe II using the optical follow-up observations. They are an explosion of an extended progenitor star, and thus their slow mass-loss wind velocity ($v_{\rm w} \sim 10$ km s$^{-1}$) limits the investigation of the mass-loss history in the final $\sim 30$ years. A similar approach for SESNe, especially SNe Ib/c from a compact progenitor star, should allow us to obtain the information on the pre-SN activity in the truly final period within the last year, thanks to the expected high mass-loss wind velocity for these compact progenitors. However, it is expected that the optical data are not sensitive to the properties of the CSM for (canonical) SESNe. An alternative approach is required, and we demonstrate in the present work that the quick radio follow-up observations, in particular those at high-frequency data, provide a powerful tool to overcome the difficulty and go beyond the present boundary in investigating the final evolution of massive stars. 

We have presented our ALMA band 3 (at 100 GHz) observations of a nearby SN Ic 2020oi. Through qualitative (and largely model-independent) analyses and quantitative light curve model calculations performed on the whole radio data set as combined with the lower-frequency data \citep{horesh2020}, we have reconstructed the radial CSM distribution from the very vicinity of the SN ($\lsim 10^{15}$ cm) to the outer region (a few $\times 10^{16}$ cm), which traces the progenitor activity in its final phase, down to the sub-year time scale toward the SN explosion (for the assumed mass-loss velocity of $v_{\rm w} = 1,000$ km s$^{-1}$). 

We have argued that the CSM structure, derived under the standard assumptions on the SN-CSM interaction and the synchrotron emission, shows deviation from a smooth distribution expected from the steady-state mass-loss. The radio properties (the SED and the decay slope) of SN 2020oi in the late-phase ($\gsim 40$ days) are quite typical of SESNe; the smooth CSM structure is roughly approximated by the steady-state wind. However, in the earlier phase ($\lsim 40$ days), the radio properties are not explained by simply extending the smooth CSM to the inner region. The main argument is the temporal evolution covering the IC cooling phase. The smooth CSM distribution (described by a single power law) predicts that the (optically-thin) light curve decays steeply toward the optical peak, then the light curves should be flattened after the peak, then it will eventually become steeper again to reach to the adiabatic regime; namely, the steep-flat-steep evolution is generally expected in a model independent way. The observed behavior is however opposite; the flat-steep-flat evolution is seen, where the ALMA data play a critical role to identify the flat evolution in the earliest phase. This indicates that the CSM radial distribution follows the flat-steep-flat density distribution from the inner to the outer regions. 

The existence of a moderately dense component is thus derived in the very vicinity of the SN ($\lsim 10^{15}$ cm). Being an explosion of a compact C+O star having a fast wind, it indicates that the mass-loss property shows fluctuations on the sub-year time scale. This finding suggests that the pre-SN activity is likely driven by the accelerated change in the nuclear burning stage in the last moments just before the massive star's demise. 

The structure of the CSM derived in this study is beyond the applicability of the other methods at optical wavelengths, highlighting an importance and uniqueness of quick follow-up observations of SNe by ALMA and other radio facilities. The present study provides a proof-of-concept for such investigation, and we plan to apply similar analyses for a sample of not only SESNe but also SNe II. The particular example presented in this paper indicates that the mass-loss rate derived for SN Ic 2020oi in the last few years is likely at a similar level derived for SNe II in the last few decades. This may indicate that a common property is likely shared by the final evolution of the progenitors of SESNe and SNe II. The modest fluctuation in the final sub-year time scale may also be common, which might be simply missed for SNe II due to the observational difficulty. 

However, since the nature of the progenitor stars for SESNe can be diverse, it is required to perform the similar analyses for a sample of objects. 
Indeed, SN Ic 2020oi belongs to the population showing the fastest evolution among SESNe \citep{horesh2020,rho2021,ho2021}, and it might originate from a progenitor star whose mass is around the lowest boundary to become an SN Ic \citep{gag2021}. The previous example of the early radio follow-up data for SN Ib iPTF13bvn indicates that the mass-loss rate corresponding to the innermost CSM distribution (up to a few $10^{15}$ cm) would be only a few $\times 10^{-5} M_\odot$ yr$^{-1}$ \citep{cao2013}, which is about an order of magnitude lower than that derived for SN 2020oi in the present work; unfortunately the data are only available for the early phase up to $\sim 10$ days for iPTF 13bvn, and thus it is not clear whether iPTF 13bvn also had the non-smooth CSM distribution. Further, a fair comparison between the innermost CSM densities for the two cases will require a careful analysis and modeling under the same model formalism and the treatment of the microphysics parameters. Such investigation, not only expanding the sample of the quick and well-sampled radio data but also the systematic model analysis, will allow us to understand the nature of massive stars' evolution in their final phase. 

\acknowledgments

This paper makes use of the following ALMA data: ADS/JAO.ALMA \#2019.1.00350.T. The following ALMA data are also used as supplementary information: \#2013.1.00634S and \#2015.1.00978S. ALMA is a partnership of ESO (representing its member states), NSF (USA) and NINS (Japan), together with NRC (Canada), MOST and ASIAA (Taiwan), and KASI (Republic of Korea), in cooperation with the Republic of Chile. The Joint ALMA Observatory is operated by ESO, AUI/NRAO and NAOJ. K.M. acknowledges support from the Japan Society for the Promotion of Science (JSPS) KAKENHI grant JP18H05223 and JP20H04737. K. M. and T. J. M. acknowledge support from the JSPS KAKENHI grant  JP20H00174. P.C. acknowledges support from the Department of Science and Technology via SwaranaJayanti Fellowship award (File no.DST/SJF/PSA-01/2014-15). P. C. also acknowledges support from Department of Atomic Energy, government of India, under the project no. 12-R\&D-TFR-5.02-0700. T. M. acknowledges support from JSPS KAKENHI grant 21J12145 for the fiscal year 2021 and from the Iwadare Scholarship Foundation for the fiscal year 2020. H.K. was funded by the Academy of Finland projects 324504 and 328898.

%






\bibliography{sn2020oi_alma}{}
\bibliographystyle{aasjournal}



\end{document}